\documentclass{ws-ijmpa}

\begin{document}

\markboth{Kim Maltman}{Heavy Antiquark Pentaquarks in the CM and GB Models}

\catchline{}{}{}{}{}

\title{Heavy Antiquark Pentaquarks in the CM and GB Models}

\author{Kim Maltman}
\address{Dept. Math. and Stats, York Univ., Toronto, ON CANADA\\
CSSM, Univ. of Adelaide, Adelaide, SA, Australia}


\maketitle


\begin{abstract}
The splittings between positive parity pentaquarks containing a single
heavy ($c,b$) antiquark and four light ($u,d$) quarks
are investigated in models having spin-dependent
interactions generated by either effective Goldstone boson exchange
or effective color magnetic exchange.
Model-independent features of these
splittings are emphasized. Overlaps to the ``fall-apart''
decay configurations $ND$, $ND^*$ or $NB$, $NB^*$,
relevant to decay couplings for the various 2-body 
pentaquark decays, are also computed.
\end{abstract}



\section{Introduction}

Recent experimental observations of the $S=+1$ $\theta$ baryon\cite{thetaexp},
together with reports of signals for an $I=3/2$ $\Xi$ state\cite{NA49}
and a $C=-1$ baryon state\cite{H1} have created intense theoretical
interest in pentaquark states. Both chiral soliton\cite{chsm} and 
quark model\cite{gbposparheavy,jw,kl,jm,qm,kmheavy}
scenarios for the nature of these states have been investigated,
but to date no clear theoretical consensus has emerged.

A key goal for model-dependent approaches is to find predictions 
unaffected by uncertainties in either model parameters or 
incompletely understood dynamical model features. For quark model
approaches based on effective color magnetic (CM) or Goldstone boson 
(GB) exchange, we show that the pattern of splittings in the positive 
parity ($P=+$) $\bar{Q}\ell^4$ ($Q=c,b$, $\ell =u,d$) sectors provides
several such predictions, allowing the models to be tested
(and potentially ruled out) by future experiments.
We concentrate on the $P=+$ sector, corresponding to
relative p-wave meson-baryon decays, since, due to
the centrifugal barrier, attractive short-range interactions 
insufficiently strong to bind can still produce resonant behavior.
We study {\it all} Pauli allowed states of the form $\bar{Q}\ell^4$ having
light quark $[31]_L$ orbital and $[211]_c$ color symmetry. 
(The optimal hyperfine attraction is much greater in
the $[31]_L$ than in the $[4]_L$ orbital sector~\cite{gbposparheavy,jm}.)
For the explicit forms of the spatial
wavefunctions, see Ref.~\cite{kmheavy}.
For the GB case, the spin-dependent interactions, $H_{GB}$,
depend on spin and flavor (see Ref.~\cite{gbposparheavy} for details);
for the CM case they depend on color and spin (see Ref.~\cite{kmheavy}
for details of $H_{CM}$).

For each $(I,J_q)$ (with $J_q$ the total intrinisic spin),
$H_{GB,CM}$ are diagonalized in the basis of all Pauli-allowed states. 
The collection of these results yields the ground state quantum numbers 
and excitation spectrum. The {\it pattern} of splittings
is determined solely by the structure of $H_{GB,CM}$; the overall size
scales with that of the spatial matrix element
of a relative s-wave $\ell\ell$ pair (see Ref.~\cite{kmheavy} for details).

\section{Relation to the Jaffe-Wilczek and Karliner-Lipkin Scenarios}

In the Jaffe-Wilczek (JW) scenario~\cite{jw}, the $\theta$
is assumed to consist of two $I=J_q=0$, $C=\bar{3}$ $qq$ pairs coupled
to $C=3$, with only confinement acting
between the $\bar{s}$ and $qq$ pairs. Cross-cluster
antisymmetrization (AS'n) and interaction effects are assumed 
suppressed by the relative p-wave between the diquark pairs.
The same $\ell^4$ configuration is expected for the 
heavy pentaquark ground states, $\theta_{c,b}$.
The analogous baryon splittings,
$m_{\Lambda_{c,b}}-m_\Lambda$,
are used to estimate $m_{\theta_{c,b}}-m_\theta$, yielding
\begin{equation}
\left[ m_{\theta_c}\right]_{JW}\simeq 2710\ {\rm MeV}, \quad
\left[ m_{\theta_b}\right]_{JW}\simeq 6050\ {\rm MeV}\ ,
\label{jwthetac}\end{equation} 
$\sim 100$, $170$ MeV, respectively, below $N\bar{D}$, $NB$ thresholds. 

In the GB model, 
the JW correlation is by far the most attractive $qq$ correlation.
The model, with GB exchange acting only in the $\ell^4$ sector,
thus corresponds to the JW scenario and provides
a dynamical framework for studying (1) corrections 
associated with inclusion of cross-cluster 
AS'n/interaction effects and/or configuration mixing,
and (2) the pattern of
excitations above a JW-like ground state, not
addressed in the JW scenario. The model yields JW ground state
quantum numbers, a $\theta_{c,b}$ $\ell^4$ configuration identical to that 
in the $\theta$, but a ground state hyperfine expectation $\sim 35\%$ more 
attractive than the strict JW scenario value.

In the (CM-motivated) Karliner-Lipkin (KL) scenario, 
the $\theta$ consists of one $I=J_q=0$, $C=\bar{3}$ diquark and
one $(I,J_q,C)=(0,{\frac{1}{2}},3)$ ``triquark'' (in which
the spin and color of a $J_q=1$, $C=6$ $ud$ pair are
anti-aligned to those of the $\bar{s}$)~\cite{kl}. The 
KL correlation has lower CM hyperfine energy than does the 
JW correlation. The $q\bar{Q}$
interactions (responsible for favoring the triquark configuration)
are weakened when $\bar{s}\rightarrow \bar{c},\bar{b}$,
reducing the hyperfine attraction for the $\theta_{c,b}$.
KL estimate this reduction by assuming that (i) the $\theta_{c,b}$ and 
$\theta$ share the same diquark-triquark structure, and (ii) the
$\bar{Q}$ hyperfine interactions scale, as in
the CM model, with $1/m_{\bar{Q}}$~\cite{kl}. The 
resulting modified JW-style estimates are
$\sim 180$ MeV above strong decay thresholds:
\begin{equation}
\left[ m_{\theta_c}\right]_{JW}\simeq 2985\ {\rm MeV}, \quad
\left[ m_{\theta_b}\right]_{JW}\simeq 6400\ {\rm MeV}\ .
\label{klthetac}\end{equation} 

In the $\theta$ sector of the CM model, however, the
same $\ell\bar{s}$ interactions which lower 
the KL triquark energy also mix the JW and KL 
configurations~\cite{jm} (an effect also weakened
when $\bar{s}\rightarrow \bar{c},\bar{b}$).
The $I=0,J_q=1/2$ ground state hyperfine expectation is significantly
lower than either the JW or KL expectations, and well approximated 
using the optimized combination of JW and KL correlations~\cite{jm}. 
For conventional values of $m_{\bar{b},\bar{c}}$,
the KL correlation is less attractive that the JW correlation 
in both the $\theta_{c,b}$ systems; hence the JW, and not the KL,
correlation is expected to dominate these systems. 
The CM model ground state hyperfine
expectation, including cross-cluster AS'n/interaction
effects neglected by KL, is, indeed, within $\sim 13\%$ (though 
less attractive) of the JW correlation estimate. The resulting
modified JW-style $\theta_{c,b}$ mass estimates
lie just above strong decay threshold for the $\theta_c$,
and just below for the $\theta_b$:
\begin{equation}
m_{\theta_c}\simeq 2835\pm 30\ {\rm MeV}, \quad
m_{\theta_b}\simeq 6180\pm 30\ {\rm MeV}\ .
\label{modifiedthetac}\end{equation}


\section{Results and Conclusions}

Results for the splittings and overlaps are given in Tables 1, 2.
Column 1 lists $(I,J_q)$,
column 2 $\Delta\hat{E}$, the hyperfine splitting relative to
the $(I,J_q)=(0,1/2)$ ground state, in dimensionless units~\cite{kmheavy}.
The scale of the splittings in physical
units is obtained by restoring an overall factor
involving the hyperfine expectation
for a relative s-wave $\ell\ell$ pair~\cite{kmheavy}. 
Estimating this factor using the corresponding
$N$ expectation (equivalently, using the $\Delta$-$N$ splitting)
yields the estimated physical splitting values, $\Delta E^{est}$,
given in Column 3.
Columns 4 and 5 contain,
for each excited pentaquark state $P^*$, 
the squares of the ratios $g_P$ and $g_{V^*}$,
defined by
\begin{equation}
g_P\, =\, \langle NP_H\vert P^*\rangle 
/ \langle NP_H\vert P_{gnd}\rangle , \quad
g_{V^*}\, =\, \langle NV^*_H\vert P^*\rangle
/ \langle NP_H\vert P_{gnd}\rangle \ ,
\label{gdefns}\end{equation}
with $P_{gnd}$ the corresponding $(I,J_q)=(0,1/2)$ ground state,
and $P_H$, $V^*_H$ the corresponding heavy pseudoscalar and vector mesons.
The relative coupling strengths 
for the decays $P^*\rightarrow NM$, with $M=P_H, V_H^*$, 
should be given by the ratios of these factors if the dominant 
mechanism for $P=+$ pentaquark decay to $NM$ 
is ``fall-apart'' through the p-wave barrier~\cite{closezhao}.

\begin{table}[h]
\tbl{Low-lying positive parity excitations of the $\theta_{c,b}$ in the
GB model.}
{\begin{tabular}{@{}lcccc@{}} \toprule
$\ (I,J_q)$&$\Delta\hat{E}$&$\Delta E^{est}$ (MeV)&$g_P^2$&$g_{V^*}^2$\\
\hline
(0,1/2)&0&0&1&3.00\\
(1,1/2)&4.50$\rightarrow$5.71&132$\rightarrow$167&2.24$\rightarrow$2.54
&0.75$\rightarrow$0.85\\
(1,3/2)&4.50$\rightarrow$5.71&132$\rightarrow$167&0
&1.27$\rightarrow$1.36\\
(0,1/2)&10.2$\rightarrow$14.5&299$\rightarrow$423&2.01$\rightarrow$2.07
&0.67$\rightarrow$0.69\\
(0,3/2)&10.2$\rightarrow$14.5&299$\rightarrow$423&0
&2.68$\rightarrow$2.75\\
\botrule
\end{tabular}}
\end{table}

\begin{table}[h]
\tbl{Low-lying positive parity excitations of the $\theta_{c,b}$ in the
CM model.}
{\begin{tabular}{@{}llcccc@{}} \toprule

Sector&$\ (I,J_q)$&$\Delta\hat{E}$&$\Delta E^{est}$ (MeV)&$g_P^2$&$g_{V^*}^2$\\
\hline
\hline
Charm&(0,1/2)&0&0&1&0.74$\rightarrow$2.22\\
&(0,1/2)&1.14$\rightarrow$1.20&84$\rightarrow$88&0.55$\rightarrow$1.87
&1.54$\rightarrow$2.32\\
&(1,1/2)&1.22$\rightarrow$1.47&89$\rightarrow$108&1.95$\rightarrow$3.41
&0.03$\rightarrow$0.35\\
&(0,3/2)&1.29$\rightarrow$1.56&94$\rightarrow$114&0
&1.60$\rightarrow$2.79\\
&(1,3/2)&1.61$\rightarrow$1.87&118$\rightarrow$137&0
&0.85$\rightarrow$1.52\\
&(1,1/2)&1.79$\rightarrow$2.07&131$\rightarrow$152&0.00$\rightarrow$0.14
&1.72$\rightarrow$2.72\\
\hline
\hline
Bottom&(0,1/2)&0&0&1.00&1.87$\rightarrow$2.71\\
&(0,1/2)&1.16$\rightarrow$1.25&85$\rightarrow$92&1.54$\rightarrow$2.32
&0.88$\rightarrow$0.94\\
&(0,3/2)&1.26$\rightarrow$1.35&92$\rightarrow$99&0
&2.51$\rightarrow$3.21\\
&(1,1/2)&1.43$\rightarrow$1.55&105$\rightarrow$114&1.76$\rightarrow$3.65
&0.20$\rightarrow$0.76\\
&(1,3/2)&1.58$\rightarrow$1.66&116$\rightarrow$122&0
&1.36$\rightarrow$1.76\\
&(1,1/2)&1.77$\rightarrow$1.99&130$\rightarrow$146&0.05$\rightarrow$0.46
&2.53$\rightarrow$2.76\\
\botrule
\end{tabular}}
\end{table}


The tables (and extended versions
thereof~\cite{kmheavy}) show the following:
(1) the $P=+$ ground state has $I=J_q=0$ for both the CM and GB models;
(2) the lowest excitation above this state is considerably lower in the CM
than in the GB model;
(3) the spectrum of excitations is much denser in the CM than
in the GB model, 5 $C=-1$ baryons being predicted in the interval between
$\sim 90$ and $130$ MeV (respectively between $\sim 150$ and $330$ MeV)
above the lowest state in the CM (respectively GB) case;
(4) for the ``low-lying'' excitations just noted, there are
no large deviations from $1$ in the relative two-body decay couplings;
(5) five additional excitations having rather small couplings
to the two-body decay channels are also expected within
$\sim m_\Delta -m_N$ of the ground state in the CM model.


\section*{Acknowledgements}
The support of the Natural Sciences and Engineering Research Council of Canada
is gratefully acknowledged.


\end{document}